\documentclass[reqno,12 pt,a4]{amsart}
\usepackage[centertags]{amsmath}
\usepackage{amsfonts}
\usepackage{amssymb}
\usepackage{amsthm}
\usepackage{newlfont}
\usepackage{epsfig}

\theoremstyle{plain}

\theoremstyle{definition}

\theoremstyle{remark}

\numberwithin{equation}{section}

\newcommand{\opunit}{\text{1}\kern-0.22em\text{l}}

\newcommand{\half}{\mbox{\tiny $\frac12$}}

\begin{document}

\begin{center}

\noindent{\large \bf Thermoelectric phenomena via \\an interacting particle system } \\

\vspace{15pt}

{\bf Christian Maes} and {\bf Maarten H. van Wieren}\footnote{Corresponding author:
{\tt Maarten.vanWieren@fys.kuleuven.ac.be}}\\
Instituut voor Theoretische Fysica\\ K.U.Leuven, Belgium.

\end{center}

\vspace{20pt} \footnotesize \noindent {\bf Abstract: } We present
a mesoscopic model for thermoelectric phenomena in terms of an
interacting particle system, a lattice electron gas dynamics that
is a suitable extension of the standard simple exclusion process.
We concentrate on electronic heat and charge transport in
different but connected metallic substances. The electrons hop
 between energy-cells located alongside the spatial extension
of the metal wire. When changing energy level, the system
exchanges energy with the environment. At equilibrium the
distribution satisfies the Fermi-Dirac occupation-law. Installing
different temperatures at two connections induces an electromotive
force (Seebeck effect) and upon forcing an electric current, an
additional heat flow is produced at the junctions (Peltier heat).
We derive the linear response behavior relating the Seebeck and
Peltier coefficients as an application of Onsager reciprocity. We
also indicate the higher order corrections.  The entropy
production is characterized as the anti-symmetric part under
time-reversal of the space-time Lagrangian.

\vspace{5pt}  \footnotesize \noindent {\bf KEY WORDS:} interacting
particle system, Seebeck and Peltier effect, fluctuations and
response functions.

\normalsize
\section{Introduction}
The goal and motivation of the paper comes form exploring
nonequilibrium behavior via the theory of spatially extended
stochastic dynamics, a standard reference is \cite{Sp}.
We present an interacting particle system for the standard
thermoelectric phenomena such as the Seebeck and the Peltier effect.
Second, we show how the steady state fluctuation theorem rigorously
reproduces the phenomenological linear response theory. Finally,
the model also provides a testing ground for deriving higher-order
response functions and for exploring the nonequilibrium theory beyond
linearity.
\\

Thermoelectric systems provide a phenomenology for the foundations
of nonequilibrium thermodynamics.  Standard examples of
thermoelectric phenomena are the Peltier effect (1834) ---
electric current through different metallic conductors will in
general cause production or absorption of heat at the junctions,
and the Seebeck effect (1826) --- an e.m.f. (electro-motoric or thermoelectric force)
will appear in the circuit if the junctions are maintained at a
different temperature (thermocouple). They are discussed in the
standard textbooks such as \cite{deGM} and except for a short
reminder in Section \ref{phen} we assume the reader familiar with
the basic facts.\\
 As written in the first line of Onsager's 1931
article on reciprocal relations,``When two or more irreversible
transport processes (heat conduction, electrical conduction and
diffusion) take place simultaneously in a thermodynamic system the
processes may interfere with each other.''  Reciprocity relations
are the most well-know realizations of that and Thomson was
probably the first (in 1854) to propose them.  Onsager took up the
derivation of the reciprocity relations now named
after him \cite{ons}. In the second half of the previous century the linear
response theory added the response relations such as these of
Green and Kubo from which the Onsager relations are explicitly
visible.\\
  The more recent history has seen a revival in
attacking questions on the construction of nonequilibrium
statistical mechanics.  One proposed generalization of (linear)
fluctuation-dissipation relations has been the so called
fluctuation theorem, both in its transient and in its steady state
version. For steady states, it gives a symmetry in the
distribution of the variable entropy production.  We  derive it
here in Section \ref{ep2}. Section \ref{sec:Lin resp} explains how
it leads to the phenomenological relations between Seebeck and
Peltier coefficients.  Not emphasized in the literature so far is
that this fluctuation symmetry is useless for higher order
response functions; it is unable to produce a useful relation for
the second order derivatives of the current around equilibrium. We
observe in Section \ref{sec:gen_funct_gen_obs} how to remedy the
situation for our particular model.

The physical origin of the model is discussed in Section \ref{pi}.
From the point of view of the  standard theory of interacting
particle systems, it is an adaptation of exclusion processes, see
\cite{KL} and \cite{lig} for mathematical background. In that way it is easy to
simulate and it thus offers an environment for studying
analytically less accessible aspects of thermoelectric phenomena
(such as the influence of disorder). From the point of view of
solid state physics, it can be seen as the result of
time-dependent perturbation theory to transport problems in
metals.  We only keep the electron hopping with transition
probabilities that satisfy a local detailed
balance condition.\\
  The details of the model are explained in
Section \ref{genmod} while, as a warm-up, Section \ref{heath}
contains its most simplified form.

\section{Toy model for heat conduction}\label{heath}\label{sec:toy
model} Consider a lattice rectangle of height 2 and width $M$, see
Fig.1. Each of the  $2M$ sites can be occupied by at most one
particle.  Think of the rectangle as a `metal wire' of size $M$ in
which each couple
 of sites $(x,2M+1-x)$ forms one cell with two energy
levels
 that can each be occupied by one electron. The energy
difference between the two levels is taken equal to a constant $e>0$.\\
 We consider a Markov process on the state space
$K=\{0,1\}^{2M}$ in which particles can hop on nearest neighbor
sites.  The hopping is symmetric (subject only to the simple
exclusion rule of one particle per site) except for the bonds
$(2M,1)$ and $(M,M+1)$. More specifically, an elementary
transition consists of randomly picking a site $x$ and changing
the configuration $\eta$ into the new configuration $T_x\eta$
which remains equal to $\eta$ on all sites $y$ except for $y=x,
T_x\eta(x) =\eta(x+1)$ and for $y=x+1, T_x\eta(x+1)=\eta(x)$.  The
transition rate for $\eta \rightarrow T_x\eta$ is equal to
\begin{eqnarray}
\exp[\beta_A\,e\,(\eta(x+1)-\eta(x))/2],&&\quad \mbox{ if } x=2M\nonumber\\
\exp[\beta_B\,e\,(\eta(x)-\eta(x+1))/2],&&\quad \mbox{ if } x=M\nonumber\\
1 ,&&\quad \mbox{ otherwise }
\end{eqnarray}
   The number of particles is
conserved in the evolution and the $\beta_A$ and $\beta_B$ are the
only (other) parameters in the model. In words, the particles jump
to nearest neighbor cells but remain on the same energy level (no
effective scattering in the bulk). In the left and right end
cells, particles undergo thermal transitions; they can change
energy level at inverse temperatures $\beta_A$ and $\beta_B$
respectively.  The system is thus at its both ends in contact with
a heat reservoir and one expects a heat current flowing whenever
they have unequal temperatures.

[FIGURE 1]

The model satisfies the condition of detailed balance when left
and right $\beta_A=\beta_B=\beta$:  For each function $F$ of the
total particle number, the probability
\[
\rho_{\beta}(\eta) = \frac{1}{Z}\,F(\sum_{x=1}^{2M}
\eta(x))\,\exp[-\beta e\sum_{x=1}^M \eta(x)], \quad \eta\in K
\]
is  stationary and reversible for the dynamics. These equilibrium
weights are completely determined by the number of particles
$\sum_{x=1}^M \eta(x)$ in the upper energy level. In order to make
our calculations possible we make use of the equivalence of
ensembles to express the single site marginal in terms of chemical
potentials. In equilibrium the grand canonical setup is made by
attaching particle reservoirs with chemical potential
$\mu_A=\mu_B=\mu$ to the left and right side of the system. For
this setup we now make the choice $F(n) = \exp[\beta\mu n]$;
$\rho_\beta$ is a homogeneous product measure with single site
marginal
\begin{equation}\label{FDdis}
n_\beta(x) \equiv \mbox{Prob}_{\rho_\beta}[\eta(x)=1] =
\frac{1}{e^{\beta [e(x) -\mu]} + 1}
\end{equation}
where $e(x)=0$ or $e$ depending on whether the $x$ is `down' $M+1
\leq x \leq 2M$ or `up' $1\leq x\leq M$ in the energy cell.
\\
 If the reservoir temperatures are unequal, when
$\beta_A\neq \beta_B$, there is no explicit expression for the
stationary measures but we do know the density profile $n(x)
\equiv $ Prob$_\rho[\eta(x)=1]$ in a stationary measure $\rho$. In
the bulk and because of the symmetric exclusion rule, there is a
linear density profile in both the upper and the lower energy
layer:
\begin{eqnarray}\nonumber
 n(x)= n^A_u+\frac{x-1}{M-1}[n^B_u- n^A_u]
 && 1\leq x\leq M
 \\\nonumber
n(x)= n_d^A+ \frac{2M-x}{M-1}[n^B_d-n^A_d]
 && M+1\leq x\leq 2M
\end{eqnarray}
but the density in each cell (containing the couple $(x,2M+1-x),
1\leq x\leq M$) is constant for a canonical system due to particle
conservation: $n(x) + n(2M+1-x) = n^B_u + n^B_d = n^A_u + n^A_d$.
In the steady state, there is no particle current between cells
but there may be a heat current.

On the upper energy layer particles transport an energy equal to
$e$; giving rise to a steady energy current, taken negative in the
$x-$direction since we will heat up side $B$, equal to
\begin{equation}\label{heatcur}
\langle J_Q\rangle= e\,\frac{n^B_u-n^A_u}{M-1}
\end{equation}
Remember that $n^A_u$ and $n^B_u$ are steady densities (left and
right on the upper energy layer depending on $M, \beta_A$ and
$\beta_B$. On the other hand, as $M\uparrow +\infty$ (very long
wire) these densities tend to their respective equilibrium values
as for example from \eqref{FDdis},
\begin{eqnarray}\label{eqvalues}
  n^A_u \rightarrow  \frac{1}{\exp(\beta_A [e -\mu_A]) + 1} &&
  n_u^B \rightarrow  \frac{1}{\exp(\beta_B [e -\mu_B]) + 1}
\end{eqnarray}
That is an application of convergence results for a sequence of
Markov processes, see e.g. Chapter 4.9 in \cite{EK}. It should be
emphasized that $\mu_A$ and $\mu_B$ are a function of $\beta_A$
and $\beta_B$ respectively, and of course as well of the particle
density. An expression for $\mu_A$ and $\mu_B$ may be derived by
solving the following identity:
\[
\rho=\half(n_u^B+n_d^B)=\frac{\half}{\exp(\beta_B [e -\mu_B]) +
1}+\frac{\half}{\exp(\beta_B [ -\mu_B]) + 1}
\]
Naturally the identity for side $A$ is similar.\\
As a result of \eqref{eqvalues}, the heat current \eqref{heatcur}
satisfies the stationary Fourier law with
\begin{equation}\label{eq:kappa=delta rho}
0 < \kappa \equiv  \lim_M M \langle J_Q\rangle < +\infty
\end{equation}
Writing out the right-hand side of \eqref{heatcur} with the
substitution \eqref{eqvalues} for $\beta_A = 1/T$ and
$\beta_B=1/(T+\epsilon)$, we get a heat conductivity equal to
\[
\lambda \equiv  \lim_{\epsilon\rightarrow 0}
\frac{\kappa\,T}{\epsilon} =  \mu^2H(0)+ (\mu-e)^2H(e)+ \frac
{(\mu H(0)+(\mu-e) H(e))^2}{H(0)+H(e)}
\]
with $H(e)=1/4\cosh^2(\beta(\mu-e))$. The last term is the
contribution due to the canonical ensemble whereas the first two
terms would also occur in a derivation for a grand canonical
setup.
\\
We conclude that this toy-model already shows electronic heat
conductivity in a clear and simple way.\\

The next Section extends the model by allowing multiple {\it
upper} and {\it lower} energy levels and by interpreting the {\it
upper} and {\it lower} levels as characterizing two different but
connected metal wires, see Fig.2.  Again  at the two junctions
there may be different temperatures and we will also consider the
case where electric power is applied.

\section{The general model}\label{genmod}
\subsection{Physical input}\label{pi}
Metals are excellent conductors of both heat and electricity.
Already the empirical law of Wiedemann and Franz (1835) relating
thermal and electrical conductivities, points to the electrons as
heat carriers.  While say for the specific heat of a metal at room
temperature there is no observable electronic contribution, for
nonequilibrium purposes mostly electrons are  responsible for heat
and charge transport. As the electron mass is small, even at room
temperatures and for typical metallic electron densities, the
Maxwell-Boltzmann and Fermi-Dirac distributions are very
different. Therefore, the Pauli exclusion principle must play a
role and the quantization into energy levels, even if only near
the Fermi energy, must be part of any physical model of transport
in metals.\\
 On the other hand a fully quantum mechanical and
microscopic treatment of transport problems is typically out of
reach.  We have here a system with extremely many degrees of
freedom and one usually finds refuge in perturbation theory.  The
electrons are moving almost independently but subject to (weak)
interactions with phonons, impurities and imperfections.  When the
average time between collisions multiplied with the Fermi energy
is larger than Planck's constant (as is the case in metals),
Fermi's Golden Rule can be applied to the transition probabilities
between electronic states and in ways similar to that for the
Boltzmann equation, one
obtains a Master Equation.\\
 The resulting model that we take up
in the present paper is mesoscopic (only electrons moving on a
lattice) and while it is thus physically motivated from
time-dependent perturbation theory for quantum mechanical
processes, it fits perfectly well in the classical theory of
interacting particle systems.\\

   One imagines two wires
connected at their ends. The wires (long thin bars) consist of a
large array of cells, discrete lattice points, at which are
located a finite number of energy levels.  Free
electrons are hopping between nearest neighbor cells, subject to an exclusion rule
and with transition rates that are specified via a local detailed
balance condition.
\\
We start by introducing the general set-up. The simplest version
was already in the previous section and other simplifications
follow later, see Section \ref{sv}.

\subsection{State space}
Consider a ring of $N$ points or cells. The situation one should
keep in mind is that of connecting two different wires, the first
of length $M$, the second of length $N-M$. To each cell in the
first wire, $1\leq k \leq M$, is assigned a set $\cal{E}_1$ of
$n_1\geq 1$ energy levels , and to each cell in the second wire,
$M+1\leq k \leq N$, is assigned a set $\cal{E}_2$ of $n_2\geq 1$
energy levels. The set of energy levels at cell $k$ is in general
denoted by
 $\cal{E}_k$ but it should be understood that it equals
 $\cal{E}_1$ if $k\in \{1,\ldots,M\}$
and that  $\cal{E}_k = \cal{E}_2$ for $k\in \{M+1,\ldots,N\}$.  A general
element of $\cal{E}_k$ is written as $e,e',\ldots$.
These are the
electronic states (we ignore spin). There can be at most one
particle per energy level and we write $x=(k,e), \eta(x)=\eta(k,e)
\in \{0,1\}$ for $e\in \cal{E}_k$. The total configuration space
is thus $K =\times_{k=1}^M \{0,1\}^{n_1}\times_{k=M+1}^N
 \{0,1\}^{n_2}$ and its elements are
denoted by $\eta,\xi,\ldots$.

[FIGURE 2]

\subsection{Transitions}\label{trans}
In each of the two wires particles can hop within the same energy
level to nearest neighbor cells. That is subject to the exclusion
rule and possibly biased in one direction via some external field.
We write $E_k \geq 0$ for the electric field over the bond
$(k,k+1)$.  One can think of a gradient in chemical potential
which is locally installed but the details of the ``battery'' will
be of no concern here.  One imagines that the two wires are in a
heat bath at inverse temperature $\beta$ (Boltzman's constant
$k_B$ is set equal to one) except possibly for the
junction $(M,M+1)$ which is kept at a different temperature.\\
  Fixing
$k=1,2,\ldots,M-1,M+1,\ldots,N-1$ for the ''bulk'' bonds
$(k,k+1)$, the exchange operator associated to energy level $e\in
\cal{E}_k$ is
\begin{eqnarray}\nonumber
 (V_k^{e}\eta)(k,e) = \eta(k+1,e),&& \\\nonumber
 (V_k^{e}\eta)(k+1,e) = \eta(k,e),&& \\\nonumber
 (V_k^{e}\eta)(x) = \eta(x)&& \mbox{ otherwise}
\end{eqnarray}
and such an exchange of occupations
\[
\eta \rightarrow V_k^{e} \eta \;\mbox{ takes place at rate }\;
\exp[\beta\, E_k (\eta(k,e)-\eta(k+1,e))/2]
\]
At the junctions, $k=M$ or $k=N$, particles can also hop but now
to an arbitrary energy level
of the other wire.
These transitions are described via the exchange
operator $T_k^{e,e'}$ for $e\in \cal{E}_k, e'\in \cal{E}_{k+1}$:
\begin{eqnarray}\nonumber
 (T_k^{e,e'}\eta)(k,e) = \eta(k+1,e'),&& \\\nonumber
 (T_k^{e,e'}\eta)(k+1,e') = \eta(k,e),&& \\\nonumber
 (T_k^{e,e'}\eta)(x) = \eta(x)&& \mbox{ otherwise}
\end{eqnarray}
and
\[
\eta \rightarrow T_k^{e,e'} \eta \;\mbox{ at rate }\;
\,\exp[\beta^k(E_k + e-e')(\eta(k,e)-\eta(k+1,e'))/2]
\]
describes a local detailed balance condition.  We set
$\beta^N=\beta$ and at the other junction $\beta^M=\beta +
\Delta$. $\Delta\neq 0$ means that one junction is maintained at a
different temperature.\\
Obviously the $V_k$ are just a special case of the $T_k$.  An
immediate generalization would be to allow transitions with
$T_k^{e,e'}$ everywhere, also in the bulk, or to allow on-site
thermal exchanges. That would break bulk energy conservation but
it would allow to describe thermal exchanges with the environment
at all cells, e.g. from scattering. In the same sense, it is no
problem to add more junctions (connecting more than two wires). As
we have it now, the bath's role in the bulk is purely as a work
reservoir. One could also imagine adding particle reservoirs. By
that we mean selecting one or more cells $k$ on which at energy
level $e\in \cal{E}_k$, we have the transitions
\begin{eqnarray}\nonumber
 (A_k^e\eta)(k,e) = 1 - \eta(k,e),&& \\\nonumber
 (A_k^e\eta)(x) = \eta(x)&& \mbox {
otherwise}
\end{eqnarray}
These break particle conservation:  particles flow in or out with rates
\begin{equation}\label{ak}
 \eta \rightarrow A_k^e\eta \;\mbox{ at rate }\;
 a_k\,\exp[\beta(e-\mu_k)(\eta(k,e)-\half)]
\end{equation}
where $\mu_k$ is the chemical potential.  We will not deal with
 these generalizations more explicitly.  It is possible to include them but
 soon the notation starts to get heavy.  Let it be clear however that in no
  way do we need to introduce a temperature or chemical potential associated
   to the {\it internal} local conditions of
    the cells in the wire.  Such {\it close to equilibrium}
    conditions are mostly present in phenomenological treatments
    of thermo-electric effect.     \\

\subsection{Dynamics}
We consider a continuous time Markov process on $K$ consisting of
the two types of elementary transitions $T_k^{e,e'}$ and $V_k^{e}$
introduced above.  For fixed sets of energy levels $\cal{E}_k$ and
spatial extensions $M$ and $N$, the remaining parameters are $E_k$
(electric field at bond $(k,k+1)$), $\beta$ and $\Delta$
(inverse temperatures).\\
For $E_k>0$ work is being done on the system and
 the particles jump preferentially forward, i.e.,
in the direction $k\rightarrow k+1$.  Energy is not (necessarily) preserved in the
transition $T_k^{e,e'}$ but the total number of particles is conserved
(at least when the  $a_k=0$ in \eqref{ak}, which we will assume hereafter).\\
The backward generator $L$ of the Markov process is a sum
\begin{equation}\label{gener}
 L=\sum_{k\neq M,N}
L_k +  L_M + L_N  
\end{equation}
 with
\[
L_kf(\eta) = \sum_{e\in \cal{E}_k} \exp[\beta E_k
(\eta(k,e)-\eta(k+1,e))/2] [f(V_k^{e}\eta) - f(\eta)]
\]
\[
L_Nf(\eta) = \sum_{e\in \cal{E}_N,e'\in \cal{E}_{1}}
\exp[\beta(E_N + e-e')(\eta(N,e)-\eta(1,e'))/2] [f(T_N^{e,e'}\eta)
- f(\eta)]
\]
and
\[
L_Mf(\eta) = \sum_{e\in \cal{E}_M,e'\in \cal{E}_{M+1}}
\exp[(\beta+ \Delta)(E_M + e-e')(\eta(M,e)-\eta(M+1,e'))/2]
[f(T_M^{e,e'}\eta) - f(\eta)]
\]

Denoting the general rate as $r(\eta,\eta')$ (of course depending
on all possible parameters), we have
\begin{equation}\label{rates}
Lf(\eta) = \sum_{\eta'} r(\eta,\eta')\,[f(\eta') - f(\eta)]
\end{equation}
summarizing \eqref{gener}; the Master Equation for the
probabilities at time $t$ is as usual
\[
\frac{d}{dt}\mbox{Prob}_t(\eta)=\sum_{\eta'}\big[
r(\eta',\eta)\,\mbox{Prob}_t(\eta') -
r(\eta,\eta')\,\mbox{Prob}_t(\eta)\big]
\]
 For every initial configuration $\eta_0$ the above defines
the process $\eta_t, t\geq 0$.\\
 We are interested in a stationary
regime for fixed (but imagined very large) spatial extensions $M$
and $N$. Let $\rho$ denote a stationary measure for the above
Markov process. It could depend
 for example on the
particle density. \\
We write $P^\tau_\rho$ for the stationary process over the
time-interval $[0,\tau]$; $\omega = (\eta_t)_{t=0}^\tau$ is a path
as sampled from $P^\tau_\rho$. We will assume that there is a
finite positive constant $C$ with
\begin{equation}\label{ratios}
\frac 1{C} \leq \frac{\rho(\eta')}{\rho(\eta)} \leq C
\end{equation} for each pair $\eta,\eta'$ that can be connected
via the transitions
outlines above.\\
 In general we know very little about this
stationary process; under equilibrium conditions we can make it
more explicit.

\subsection{Equilibrium}\label{eq}
The system is in equilibrium when $\Delta=0$ and all $E_k=0$.
Consider the probability measure $\rho_{\beta,\mu}$ defined as
\begin{equation}\label{}
 \rho_{\beta,\mu}(\eta) =
 \frac1{Z}
 e^{\beta\mu\sum_x\eta(x)}
 \exp[-\beta \sum_{k=1}^N \sum_{e\in\cal{E}_k} e\,\eta(k,e)]
\end{equation}
That is a product measure but it is not (in general) homogeneous.
It is a stationary reversible measure for the equilibrium
dynamics. Its single cell marginal is given by the Fermi-Dirac
distribution, for $e\in \cal{E}_k$,
\begin{equation}\label{eq:FD distr}
 \rho_{\beta,\mu}(\eta(k,e)=1) =
 \frac{1}{e^{\beta(e-\mu)} +1}
\end{equation}
That corresponds to the grand-canonical set-up.  Except when we
add sources and sinks for the particles ($a_k\neq 0$), the total
number of particles is conserved
in the dynamics as defined above.\\
Being in steady equilibrium implies ignoring the difference
between thermodynamic past and future and on average there is no
net transport nor dissipation.

\subsection{Simplification}\label{sv}
 The simplest version is obtained by taking $\cal{E}^1 =
\{e_1\}$ and $\cal{E}^2=\{e_2\}$ singletons. There is then one
energy level per cell, $n_1=n_2=1$, and we may just as well put
$e_2-e_1=e$.  We also put $a_k=0=E_k$ for all $k$ (canonical
set-up without external field). In the bulk of each of the wires
we now only have symmetric hopping and at the two ends of the
wires (we are now talking about the bonds $(M,M+1)$ and $(N,1)$),
there are thermal transitions:
\[
\eta \rightarrow T_M \eta \;\mbox{ at rate }\; \exp[(\beta +
\Delta)\,e\,(\eta(M)-\eta(M+1))/2]
\]
through which the occupations are exchanged over the bond
$(M,M+1)$ and
\[
\eta \rightarrow T_N \eta \;\mbox{ at rate }\; \exp[-\beta
e\,(\eta(N)-\eta(1))/2]
\]
through which occupations are exchanged over the bond $(N,1)$. We
see that it formally corresponds to the toy model for heat
transport of Section \ref{heath} with $2M=N$ and $\beta_A=\beta,
\beta_B=\beta + \Delta$ at the junctions.  In that way, the model
system of Section \ref{heath} has two possible physical
interpretations, one as consisting of $M$ cells each with two
energy levels, the other as a couple of two wires each of length
$M$ and each with one energy level per cell. In that last
interpretation, the heat current in Fourier's law (defined in
\eqref{heatcur}) is in fact a particle current through the {\it
upper} wire, caused by the different temperatures $\beta_A\neq
\beta_B$ at the junctions and so provides the simplest
illustration of a Seebeck effect.  To see the Peltier effect, we
take $\Delta=0$ (equal bath temperatures) but some $E_k > 0$. A
particle current arises through the ring and energy of magnitude
$e$ gets dissipated at the junctions.



\section{Currents and dissipation}\label{curdis}

\subsection{Local particle and heat currents}\label{lcur}
  The basic physical
quantities in the phenomenon are energy and particle number.
 The electric or particle current consists
of particles hopping to nearest neighbor sites. We fix
 a path $\omega =(\eta_0,\ldots,\eta_\tau)$ in which each change (at a random time)
 corresponds to one of the transitions described under Section \ref{trans}.\\
 The net particle current (integrated
over a time-interval $[0,\tau]$) from energy level $e$ at cell $k$
to energy level $e'$ at cell $k+1$ equals
\begin{equation}\label{pcur}
J^k(e,e')(\omega) \equiv   \sum_t [\eta_t(k,e) - \eta_t(k+1,e')]
\end{equation}
for $e\in \cal{E}_k, e'\in \cal{E}_{k+1}$.  The sum in the
right-hand side of \eqref{pcur} is over all jump times $t$; the
moments at which the trajectory $\omega$ changes states. Here and
in what follows, all times in the sums are $t=t^{-}$, right before
the jump. So \eqref{pcur} is a sum of $+1$'s and $-1$'s according
to whether the particle jumps from $(k,e)$ to $(k+1,e')$ or
oppositely; the jump is counted at the time of the transition.
 When the bond $(k,k+1)$ is in the bulk of the wires ($k\neq M,N$), then
only transitions between levels of the same energy are possible
and the net number of particles jumping forward equals
\[
J^k(e,e)(\omega) = J^{e,k}(\omega) \equiv  \sum_t [\eta_t(k,e) -
\eta_t(k+1,e)]
\]
 We call
\begin{equation}\label{elcur}
J_E^k  \equiv  \sum_{e\in \cal{E}_k, e'\in \cal{E}_{k+1}}
J^k(e,e')
\end{equation}
 the variable electric current over the bond  $(k,k+1)$.\\
Secondly, energy is transported. Over the bond $(k,k+1)$ in the
bulk the time-integrated energy current equals
\[
J_H^k \equiv \sum_{e\in \cal{E}_k} e \,J^{k,e}
\]
 At each junction
$k=M,N$, there is
 a local heat exchange with
 the environment, commonly called the Peltier heat (counted positive for a
 heat transfer
  {\it into} the environment), here variable and equal to
\[
J_P^k \equiv
 \sum_{e\in \cal{E}_k,e'\in \cal{E}_{k+1}} (e-e') J^k(e,e')
\]

\subsection{Conservation laws}

In a configuration $\eta$,
\[
U_k(\eta) \equiv \sum_{e\in\cal{E}_k} e\,\eta(k,e),\;\; N_k(\eta)
\equiv \sum_{e\in\cal{E}_k} \eta(k,e)
\]
are the energy, respectively the particle number at cell $k$.
These change.  We fix again a trajectory $\omega =
(\eta_0,\ldots,\eta_\tau)$.  For $k-1,k\neq M,N$
\begin{equation}\label{bulcon}
U_k(\eta_\tau) - U_k(\eta_0) = J_H^{k-1}(\omega) - J_H^k(\omega)
\end{equation}
At the junctions $k=M,N$,
\begin{equation}\label{pelcon}
U_k(\eta_\tau) + U_{k+1}(\eta_\tau) - U_k(\eta_0)- U_{k+1}(\eta_0)
= J_H^{k-1}(\omega) - J_H^{k+1}(\omega) - J_P^k(\omega)
\end{equation}
Similar relations hold for particle conservation.  In particular,
\begin{equation}\label{parcon}
N_k(\eta_\tau) - N_k(\eta_0) = J_E^{k-1} - J_E^k
\end{equation}
As a consequence, the steady state average defines the particle
current
\[
J_E \equiv \langle J_E^k\rangle
\]
 independent of $k$.\\
In the steady state \eqref{pelcon} can be rewritten as
\[
\langle J_H^{k-1}\rangle =\langle J_P^k \rangle + \langle
J_H^{k+1}\rangle
\]
with, from \eqref{bulcon}, $\langle J_H^{k}\rangle = J_1$ constant
for $k=1,\ldots,M-1$ and $ \langle J_H^{k}\rangle = J_2$ constant
for $k=M+1,\ldots,N-1$.  Therefore, denoting the Peltier heat at
junction $(M,M+1)$ with $J_B \equiv \langle J_P^M \rangle$ and at
junction $(N,1)$ with $J_A \equiv \langle J_P^N \rangle$, we have
\begin{equation}\label{peltcon}
J_1= J_B + J_2,\;\;\;J_A + J_B = 0
\end{equation}

\subsection{Heat dissipation}
The first law of thermodynamics
\begin{equation}\label{firstl}
dU = -dQ + dW 
\end{equation}
 with
$dQ$ the dissipated heat and $dW$ the work done on the system
can be applied to our model by identifying the proper currents.
The total energy of the system
\[
U = \sum_{k=1}^N U_k
\]
is not globally conserved, see \eqref{pelcon}. The Joule heat is
the (electrical) work done
\[
W_k \equiv E_k J_E^k
\]
over the bond $(k,k+1)$. As a result, for a path $\omega
=(\eta_0,\ldots,\eta_\tau)$ the variable dissipated heat over a
time-interval $[0,\tau]$ is
\[
Q(\omega)\equiv \sum_{k=1}^N E_k \,J_E^k(\omega) +
\sum_{k=M,N}J_P^k(\omega)
\]
 Since each bond $(k,k+1)$ is assumed
attached to a large heat bath, we get that as a function of the
system's history the total change of entropy in the  environment
equals
\begin{equation}\label{entropyc}
S(\omega) \equiv \sum_{k=1}^N \beta E_k J_E^k(\omega) + \beta
J_P^N(\omega) + (\beta+\Delta)\, J_P^M
\end{equation}

\section{Entropy production}\label{ep}

\subsection{Driving forces and mean entropy production rate}

There are in fact two driving sources of nonequilibrium: one is
the temperature gradient, the others the electric field. The
temperature gradient is installed by putting different
temperatures at the junctions.  The electric field models the
presence of an external electric field or more simply, of a
battery and  the installation of an electric potential.\\
In steady state average, the steady entropy current equals the
mean entropy production
\begin{equation}
\langle S \rangle = \beta \,\sum_k E_k \langle J_E^k\rangle +
(\beta+\Delta) J_B + \beta J_A
\end{equation}
as obtained from \eqref{entropyc}.  Writing
\begin{equation}\label{drivdef}
\ss\equiv \frac 1{N} \sum_k E_k,\quad \nabla T \equiv
\frac{1}{N}\, ( \frac 1{\beta + \Delta} - \frac 1{\beta} )
\end{equation}
 and using the
conservation law \eqref{peltcon}, we have
\begin{equation}\label{mept}
\frac{1}{N}\,\langle S \rangle =   \beta \big[\ss\, J_E - \nabla T
\,J_S\big]
\end{equation}
for the entropy source strength per unit length of the wire.  The
driving forces are clearly visible: $\ss$ is the gradient of
electric potential and $\nabla T$ is the temperature gradient.
$J_S \equiv (\beta + \Delta) J_B$ is the entropy flux into the
reservoir controlling the junction at inverse temperature $\beta +
\Delta$.


The mean entropy production rate $\sigma$ over a time-interval
$[0,\tau]$ is defined as
\begin{equation}\label{mep}
\sigma_\tau\equiv  \sigma\equiv \frac 1{\tau} \int\, S(\omega)
 \,dP^\tau_\rho(\omega) = \frac 1{\tau} \langle S \rangle
\end{equation}
and can of course be rewritten from \eqref{mept}.
 We have the
fundamental result that it is positive: When not in equilibrium
$\sigma > 0$.
That can be seen as follows.  We compute
\begin{equation}\label{defr}
 R \equiv \log
\frac{dP^\tau_\rho}{dP^\tau_\rho\Theta}
\end{equation}
where $\Theta$ is the dynamical time-reversal: $\Theta\omega=(\eta_{\tau-t})_{t=0}^\tau$.
Then, by construction,
\[
\int dP^\tau_\rho(\omega)  e^{-R(\omega)} = 1
\]
and hence,
\begin{equation}\label{ster}
\int dP^\tau_\rho(\omega) R(\omega) \geq 0
\end{equation}
and is non-zero whenever $R(\omega)$ is not constant (with
$P^\tau_\rho$-probability one).  The hypothesis ``not in
equilibrium'' refers exactly to that breaking of time-reversal
invariance.  The computation of $R$ is an application of the
Girsanov formula for Markov jump processes, see e.g. the Appendix
in \cite{KL}.  It is easy to check from Section 3.4 that
\begin{equation}\label{rsrel}
R(\omega) = S(\omega) - \log \rho(\eta_\tau) + \log \rho(\eta_0)
\end{equation}
and hence the steady state expectations of $R$ and of $S$
coincide. As a result, \eqref{ster} implies the positivity of
\eqref{mep}.

\subsection{Fluctuations}\label{ep2}

Consider again \eqref{entropyc} :
\begin{equation}\label{abbr}
S = S(E,\Delta) \equiv S_0+ \sum_{k} E_k J_E^k + \Delta J_P^M,\;\;
S_0\equiv \beta(J_P^N + J_P^M)
\end{equation}
In equilibrium $E_k=\Delta=0$ and always $\langle S_0\rangle =0$.
The dissipation \eqref{abbr} is a fluctuating quantity and we can
consider its generating function
\[
F_\tau(z) \equiv \langle e^{-z S}\rangle
\]
Clearly $S\Theta=-S$ so that by combining with \eqref{defr} and
\eqref{rsrel}, we get
\[
F_\tau(z) = \int\, e^{-(1-z)S(\omega)}\,
\frac{\rho(\eta_\tau)}{\rho(\eta_0)}
 \,dP^\tau_\rho(\omega)
 \]
From assumption \eqref{ratios} and for real $z$ that implies the
fluctuation symmetry
\begin{equation}\label{star}
p(z) = p(1-z),\quad p(z) \equiv -\lim_{\tau\rightarrow +\infty}
\frac 1{\tau} \log F_\tau(z)
\end{equation}
We can however look in somewhat greater detail to the fluctuations
as they arise from the local currents.\\

Given numbers $\gamma^k(e,e')$, we write
\[
\vec{\gamma}\cdot \vec{J} \equiv \sum_{k=1}^N\sum_{e\in
\cal{E}_k,e'\in \cal{E}_{k+1}} \gamma^k(e,e')\, J^k(e,e')
\]
For example, with $\gamma^k(e,e')= \varphi^k(e,e') \equiv E_k +
(e-e')\beta^k$, $\vec{\varphi}\cdot \vec{J} = S$ the entropy
current.  If we take $\gamma^k(e,e')= \varphi_0^k(e,e') \equiv
(e-e')\beta$, then $\vec{\varphi_0}\cdot \vec{J} = S_0$ the
equilibrium value defined in \eqref{abbr}.\\
 Consider now the
following generating function for the fluctuations of the currents
\begin{equation}\label{def:p(g,p)}
 p(\vec{\gamma},\vec{\varphi})\equiv
 -\lim_{\tau\uparrow +\infty} {\frac 1{\tau}}\log \bigl[\int dP^\tau_\rho(\omega)
 e^{-\vec{\gamma} \cdot \vec{J}(\omega)}\bigr]
\end{equation}
The second argument of $p$ (the dependence on $\vec{\varphi}$)
comes of course from the dependence of the steady state law  $=
dP^\tau_\rho$ on the driving parameters.   The limit exists for
the relevant choices of $(\gamma^k(e-e'))$ via an application of
the Perron-Frobenius
 theorem, see e.g. \cite{LS}.  Moreover,
\begin{equation}\label{prop:p(g,p)=p(p-g,p)}
p(\vec{\gamma},\vec{\varphi}) = p(\vec{\varphi} - \vec\varphi_0 -
\vec{\gamma},\vec{\varphi})
\end{equation}
To see it, we start again from the definition \eqref{defr} for $R$
and observe that for all functions $f$,
\begin{equation}\label{Req}
 \int
dP_\rho^\tau(\omega) f(\Theta\omega) = \int dP_\rho^\tau(\omega)
e^{-R(\omega)}\,f(\omega)
\end{equation}
We substitute $f = \exp \vec{\gamma}\cdot \vec{J}$ and use first
that $J^k(e,e')(\Theta\omega)= -J^k(e,e')(\omega)$, the
antisymmetry under time-reversal of the variable currents. On the
other hand, $R=S-\log\rho(\eta_\tau)+\log\rho(\eta_0)$
with
\[
 S =(\vec{\varphi} - \vec{\varphi_0})  \cdot \vec{J} + S_0
\]
 given in \eqref{entropyc} or in \eqref{abbr} where, as follows
from \eqref{bulcon}--\eqref{pelcon}, the equilibrium value
\[
S_0(\omega) = \vec\varphi_0 \cdot \vec{J}(\omega) = \beta\,
[U(\eta_\tau)) - U(\eta_0)]
\]
equals the total change of energy.\\
 For that choice of $f$
\begin{eqnarray}\label{eq:lim_tau ineq}
C\,\int dP_\rho^\tau e^{-(\vec\varphi - \vec\varphi_0) \cdot
\vec{J}}\,f] &\leq &
  \int dP_\rho^\tau e^{-R}\,f
 \\\nonumber
 \leq &&
 \frac 1{C}\, \int dP_\rho^\tau e^{-(\vec\varphi - \vec\varphi_0) \cdot \vec{J}}\,f]
\end{eqnarray}
where we used again \eqref{ratios}. We conclude that after taking
the logarithm and dividing by $\tau \uparrow +\infty$, $R$ can
 be substituted with $(\vec\varphi - \vec\varphi_0) \cdot \vec{J}$
  in \eqref{Req} and the identity \eqref{prop:p(g,p)=p(p-g,p)}
 follows immediately.

\section{Linear response regime}\label{sec:Lin resp}

\subsection{Green-Kubo relations}\label{GK}

The steady state average will from now on be decorated by the
field $\vec\varphi$ to indicate its dependence on the driving:
$\langle \cdot \rangle = \langle \cdot \rangle_{\vec\varphi}$. For
the equilibrium values $\vec\varphi_0$ we write $\langle \cdot
\rangle_{\vec\varphi_0} = \langle \cdot \rangle_0$. A general
couple of energy values $e\in \cal{E}_k, e'\in \cal{E}_{k+1}$ is
denoted by $\alpha=(e,e'), J^k(e,e') = J^k_\alpha, \varphi^k(e,e')
= \varphi^k(\alpha)$. In particular, for all $\alpha$ and $k$,
\[
\langle J_\alpha^k \rangle_0 = 0
\]
is the zeroth order contribution to the steady currents.  The
Green-Kubo relations want to give details about the linear order
in $\vec\varphi$ around $\vec\varphi_0$:
\[
\langle J_\alpha^k\rangle_{\vec\varphi}=
  \sum_{\delta,\ell}
  L_{\alpha,k;\delta,\ell}\,(\varphi^\ell(\delta)-\varphi_0^\ell(\delta))
\]
up to higher order terms, with
\begin{equation}\label{lint}
L_{\alpha,k;\delta,\ell} \equiv   \frac{\partial}{\partial
\varphi^{\ell}(\delta)}\langle
J_\alpha^k\rangle_{\vec\varphi=\vec\varphi_0}
\end{equation}
linear transport coefficients.

The following Green-Kubo relation follows directly from
\eqref{prop:p(g,p)=p(p-g,p)}:
\begin{equation}\label{GKrel}
L_{\alpha,k;\delta,\ell} = \half \lim_{\tau\uparrow +\infty} \frac
1{\tau} \langle J_\alpha^k J_\delta^\ell \rangle_{0}
\end{equation}
where one should remember that the currents $J_\alpha^k$ are
extensive in $\tau$.  Deriving \eqref{GKrel}
 proceeds by differentiating the left- and
right-hand sides of \eqref{prop:p(g,p)=p(p-g,p)}.  With the
notation
\begin{equation}\label{def:C}
  C_{\alpha}^{k}g(\vec \gamma,\vec \varphi)\equiv
  \frac{\partial}{\partial \gamma_{\alpha}^{k}}g(\vec\gamma,\vec\varphi)
\end{equation}
\begin{equation}\label{def:D}
  D_{\alpha}^{k}g(\vec \gamma,\vec \varphi)\equiv
  \frac{\partial}{\partial \varphi^{k}(\alpha)}g(\vec \gamma,\vec \varphi)
\end{equation}
and under the condition that the function $p$ in
\eqref{def:p(g,p)} is smooth, we have in fact
\begin{equation}
 D_\delta^\ell C_\alpha^k p(0,0) =
 -\half C_\delta^\ell C_\alpha^k p(0,0)
\end{equation}
which implies \eqref{GKrel}. The reasoning above is a detailed
reproduction of what was observed before, e.g. in \cite{G,LS,M},
to derive fluctuation-dissipation relations from the fluctuation
symmetry \eqref{prop:p(g,p)=p(p-g,p)}. One observes that the
Green-Kubo relations \eqref{GKrel} immediately imply the Onsager
reciprocity
\begin{equation}\label{onsrec}
L_{\alpha,k;\delta,\ell} = L_{\delta,\ell;\alpha,k}
\end{equation}

\subsection{Reproducing the phenomenological behavior}\label{phen}

The above Green-Kubo relations \eqref{GKrel} are the most detailed
forms of linear response relations.  We connect them now with the
standard experimental set-up for the situation of the different
thermo-electric
effects.\\

The basic equation is \eqref{mept}.  Though we did not start from
a balance equation for the ``close to equilibrium entropy of the
system,'' \eqref{mept} reproduces exactly equation (61) in Chapter
XIII of \cite{deGM}.  The next thing is to linearize the currents
$J_S$ and $J_E$ around $\nabla T= $$\ss =0$.  For that we have the
Green-Kubo relations.  We can indeed either work from the previous
section or, more directly, use \eqref{Req} which says that
\begin{equation}\label{star1}
 \langle J \rangle
= \frac 1{2} \langle J (1 - e^{-R})\rangle
\end{equation}
for functions $J$ that are antisymmetric under time-reversal,
$J\Theta = -J$, like currents.\\
 We specify to the case where
just one of the bonds, pick $(k,k+1)$, carries an electric field
$E_k=EN$ (proportional to the length of the wire), all others
being exactly zero. That corresponds to the situation where
locally in the wire some external electric power is applied to
particles that do not interact besides exclusion. This may seem
unrealistic but it keeps things manageable and we expect the
results to be equivalent to a more realistic description. We then
have, up to linear order in $E$ and $\Delta$,
\begin{equation}
\lim_\tau \frac 1{\tau} J_S = \lim_\tau\frac 1{\tau}\big[
\frac{\beta(\beta + \Delta) EN}{2}\langle J_P^M J_E^k\rangle_0 +
\frac{\Delta(\beta + \Delta)}{2} \langle (J_P^M)^2\rangle_0\big]
\end{equation}
Rewriting that for $j_S \equiv \lim_\tau J_S/N\tau$ (per unit
length and per unit time)
\begin{equation}\label{linjs}
j_s = - L_{11}\, \nabla T  + L_{12}\, E
\end{equation}
with
\begin{eqnarray}\label{mzls1}
L_{11} \equiv && \frac{\beta}{2} \, \lim_\tau
\frac 1{\tau} \langle (J_P^M)^2\rangle_0 \nonumber\\
L_{12} \equiv && \frac{\beta(\beta + \Delta)}{2}  \lim_\tau \frac
1 {\tau} \langle J_P^M J_E^k\rangle_0
\end{eqnarray}
In the same way,
\begin{equation}\label{limlinje}
\lim_\tau \frac 1{\tau} J_E = \lim_\tau\frac 1{\tau} \big[
\frac{\beta EN}{2}\langle (J_E^k)^2\rangle_0 +
\frac{\Delta}{2}\langle J_E^k J_P^M\rangle_0\big]
\end{equation}
so that for $j_E \equiv \lim_\tau J_E/N\tau$,
\begin{equation}\label{linje}
j_E = - L_{21}\, \nabla T  + L_{22}\, E
\end{equation}
with
\begin{eqnarray}\label{mzls2}
L_{22} \equiv && \frac{\beta}{2} \, \lim_\tau
\frac 1{\tau} \langle (J_E^k)^2\rangle_0 \nonumber\\
L_{21} \equiv && \frac{\beta(\beta + \Delta)}{2}  \lim_\tau \frac
1 {\tau} \langle J_P^M J_E^k\rangle_0
\end{eqnarray}
The Onsager relation $L_{12}=L_{21}$ is apparent and hence, if we
write the standard form
\begin{eqnarray}\label{stanf}
j_S &=& -\lambda\,\beta \;\nabla T + \pi\,\beta \;j_E\nonumber\\
E &=& -v \;\nabla T + R \;j_E
\end{eqnarray}
we get $\lambda \beta \equiv L_{11} - L_{21} L_{12}/L_{22}$ for
heat conductivity $\lambda$, $R\equiv 1/L_{22}$ is a resistivity,
and the coefficients $\pi \beta \equiv L_{12}/L_{22}$ and $
v=-L_{21}/L_{22}$ verify the second Thomson relation
\[
v = -\pi \,\beta
\]
The coefficient $v$ is called the differential thermoelectric
power and $\pi \beta$ is the entropy current per electric current
when temperatures are equal, see \cite{deGM}.

\subsubsection{Seebeck effect}
That is the effect of so called thermoelectric power that can be
measured via a thermocouple.  We have  two different metals with
junctions at temperatures $T$ and $T + \epsilon$ in two heat
reservoirs. The point is that the temperature difference $\epsilon
\neq 0$ generates by itself an electro-motive force.  In other
words, we can counterbalance by a non-zero electric field $E$ and
still have zero electric current.  It can be measured in various
ways, either by inserting condensor plates in one of the metals
and seeing a potential difference or by inserting a battery in one
of the metals and measure the voltage
necessary to cancel the electric current.\\

The simplest realization in our model is again the one of Section
\ref{heath}.  Instead of thinking of one wire (with $M$ cells each
consisting of two energy levels) we can think of two different
wires (each with $M$ cells containing one energy level). What we
have called the energy current in \eqref{heatcur} now becomes the
electric or particle current. We have $\beta_A=\beta,\beta_B =
\beta + \Delta$.  We insert an electric field of strength $NE$ at
the bond $(N,1)$ (the $B-$junction). The differential
thermo-electric power or Seebeck-coefficient can now be computed
by first solving for $E$ the equation
\begin{equation}\label{solvee}
NE\,\beta_B + \beta_B \,e =  \beta\, e
 \end{equation}
 which cancels the
particle current and then putting $v = - E/\nabla T$, cf.
\eqref{stanf}. One easily finds that now
\begin{equation}\label{seec} v= - e\,\beta
\end{equation}

\subsubsection{Peltier effect}
The Peltier effect is the production/absorption of heat at the
junctions by the presence of an electric current across. Even
without establishing a temperature difference, one generates a
heat flow.  The electric current is generated by an electric field
or by an established electro-chemical gradient.  To say it
differently, to maintain a uniform and constant temperature
throughout the system, absorption of heat from the reservoir is
necessary.  That is the Peltier effect used for example to cool an
external device.\\

Let us again look at the toy model of Section \ref{heath}. By
definition,  $\pi \beta$ is the isothermal entropy flux per unit
electric current. Since the energy dissipation into the reservoir
at inverse temperature $\beta$ equals $e$ , we get $\pi \beta =
e\beta$.  Comparing with \eqref{seec}, we see again $\pi\, \beta =
-v$, the second Thomson relation.\\

In the general model particle current and heat current are not
proportional and there is no simple analogue to \eqref{solvee}.
Fortunately, the Onsager relation gives $v= -\pi\,\beta$
automatically.

\section{Beyond linear order, final remarks}\label{sec:gen_funct_gen_obs}

Despite their general appearance, fluctuation relations like
\eqref{star}, \eqref{prop:p(g,p)=p(p-g,p)} or \eqref{star1} are
not sufficient to build a systematic perturbation theory for the
physical currents beyond linear order.   The reason is that, for
say the second order term in a current, we need to know the linear
order for a time-symmetric observable (the square of the current)
and that information is absent from the fluctuation relations.
There are however alternatives.  Instead of going through general
procedures, we outline here how for the model at hand one can
compute higher order corrections to \eqref{linjs} and
\eqref{linje}.  The more systematic expansion will be part of another manuscript.\\

As is clear form Section \ref{curdis}, all steady state averages
of the form $\langle J \rangle$ (where $J$ is some current), can
be computed from the steady state average of the particle current
$J^{e,k}$ over a bond $(k,k+1)$ in the bulk. That is directly
clear from \eqref{parcon} for the electric current $J=J^k_E$ but
it also includes $J = J_P^M$ the Peltier current as we can see
from \eqref{pelcon}. We thus only need to consider one bond
$(k,k+1)$ and one energy level $e\in\cal{E}_k$ and estimate \[
\langle J^{e,k}\rangle \]
 We consider the case where we take
  a symmetric exclusion hopping process in the bulk of each line, i.e.,
  $E_k=0$ for $k\neq M,N$.  Due to the
symmetric exclusion the occupation profile on energy level $e$ is
linear and the steady state particle current on that level is
exactly given by the difference between the corresponding
occupations at the two junctions in the same sense as
\eqref{heatcur}. More precisely,  for $k=1,\ldots,M-1$
\[
 \langle J^{e,k}
\rangle = \frac{\mbox{Prob}[\eta(e,1) =1] -
\mbox{Prob}[\eta(e,M)=1]}{M-1}
\]
and similarly for the other bulk bonds.  Hence all boils down to
getting an expression for the occupations
\[
\mbox{Prob}[\eta(e,k) =1],\quad e\in \cal{E}_k, k=1,M,M+1,N
\]
To proceed we can take the same strategy as in
\eqref{heatcur}--\eqref{eq:kappa=delta rho} and suppose that $M,N$
are sufficiently large so that the occupation numbers are given by
the Fermi-Dirac equilibrium distributions. The driving parameters
are present in these occupations through the different
temperatures at the junctions and/or through the shifting of the
energy levels by the presence of a local electric field.
 We do not continue giving the details of the computation but it
should be clear that the computation is most pleasant in the
grand-canonical ensemble where the occupations, and hence all
currents can be straightforwardly computed to any given order.

\section*{Acknowledgements}
We are very grateful to Karel Neto\v cn\'y for initial conception
and conversations that started us out on the model.

\newpage
\begin{figure}
\epsfig{file=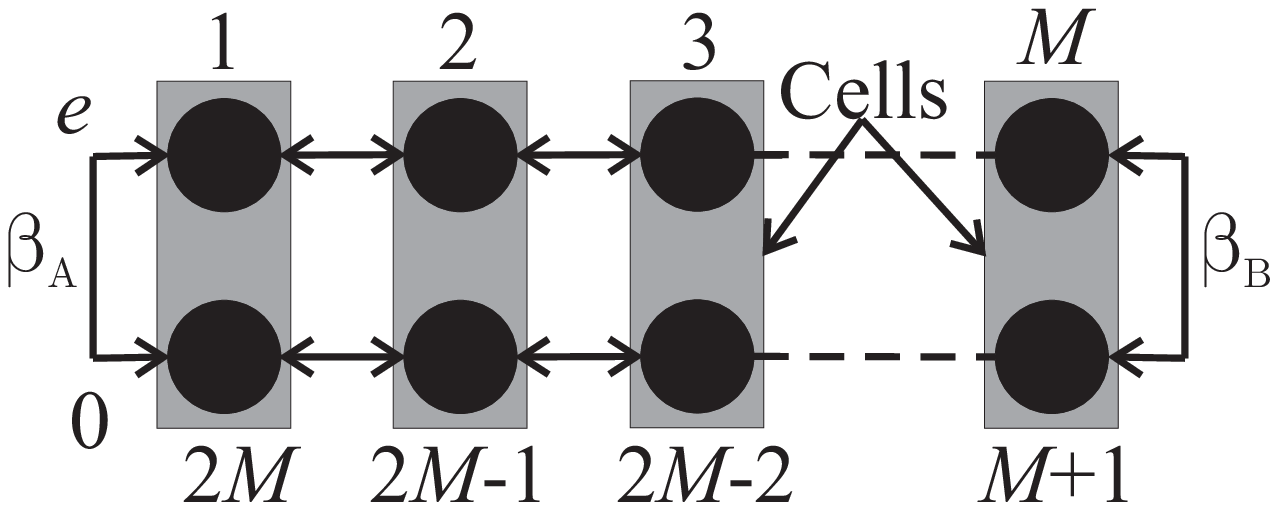,width=10cm}
\caption{\label{fig1}\footnotesize Scheme of the setup of the
toy-model.}
\end{figure}

\begin{figure}
\epsfig{file=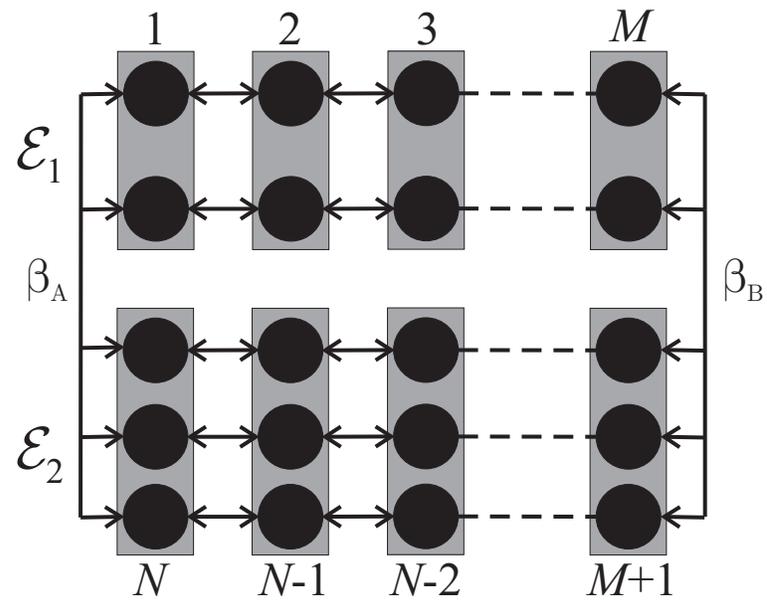,width=10cm}
\caption{\label{fig2}\footnotesize Generalized setup.}
\end{figure}

\end{document}